\documentstyle[12pt,epsfig]{article}

\newlength{\dinwidth}
\newlength{\dinmargin}
\def\upv{u^p_V}
\def\unv{u^n_V}
\def\dpv{d^p_V}
\def\dnv{d^n_V}

\def\lapproxeq{\lower .7ex\hbox{$\;\stackrel{\textstyle
<}{\sim}\;$}}
\def\gapproxeq{\lower .7ex\hbox{$\;\stackrel{\textstyle
>}{\sim}\;$}}

\def\beq{\begin{equation}}
\def\eeq{\end{equation}}
\def\bea{\begin{eqnarray}}
\def\eea{\end{eqnarray}}

\def\GeV{\rm GeV}

\begin{document}
\titlepage
\begin{flushright}
IPPP/04/62 \\
DCPT/04/124 \\
Cavendish-HEP-2004/28 \\
2nd November 2004 \\

\end{flushright}

\vspace*{0.5cm}

\begin{center}
{\Large \bf Parton
distributions incorporating QED contributions}

\vspace*{1cm}
\textsc{A.D. Martin$^a$, R.G. Roberts$^b$, W.J. Stirling$^a$
and R.S. Thorne$^{c,}$\footnote{Royal Society University Research Fellow.}} \\

\vspace*{0.5cm} $^a$ Institute for Particle Physics Phenomenology,
University of Durham, DH1 3LE, UK \\
$^b$ Rutherford Appleton Laboratory, Chilton, Didcot, Oxon, OX11 0QX, UK \\
$^c$ Cavendish Laboratory, University of Cambridge, \\ Madingley Road,
Cambridge, CB3 0HE, UK
\end{center}

\vspace*{0.5cm}

\begin{abstract}
We perform a global parton analysis of deep inelastic and related hard-scattering data,
including ${\cal O}(\alpha_{\rm QED})$ corrections
to the parton evolution. Although the quality of the fit is essentially unchanged, there
are two important physical consequences. First, the different DGLAP evolution of $u$ and $d$ type
quarks introduces isospin violation, i.e. $u^p \neq d^n$, which is found to be unambiguously in the direction
to reduce the NuTeV $\sin^2\theta_W$ anomaly. A second consequence  is the appearance of 
photon parton distributions $\gamma(x,Q^2)$ of the proton and the neutron. In principle these
can be measured at HERA via the deep inelastic scattering
processes $e N \to e\gamma X$; our predictions are in agreement with the present data.

\end{abstract}

\newpage

\section{Introduction}

Accurately determined parton distributions are an essential ingredient of precision hadron
collider phenomenology. In the context of perturbative Quantum Chromodynamics (QCD), the current frontier is next-to-next-to-leading order (NNLO), but attention has also focused recently on electroweak radiative corrections to hadron collider cross sections. Such corrections are of course routinely applied in $e^+e^-$ and $ep$ collider physics, but their application to hadron colliders is relatively new.  They have, for example, been discussed in the context of $W$ and $Z$ production
\cite{Wew,Zew} and of   $WH$ and $ZH$ production \cite{Ciccolini} at hadron colliders. 

QED contributions are invariably an important part of such electroweak corrections. 
In particular, at hadron colliders large logarithmic $\alpha \log(Q^2/m^2)$
contributions arise from photons emitted off incoming quark 
lines, the analogue of the $\alpha \log(Q^2/m_e^2)$ initial-state radiation corrections familiar
in $e^+e^-$ collisions. One could take these explicitly into account, but this would require
a consistent choice of input quark masses. Furthermore, at the very high $Q^2$ scales probed at hadron colliders, one should in principle resum these logarithms.
Fortunately the QCD factorisation theorem applies also 
to QED corrections, and as a result such collinear (photon-induced) logarithms can be 
absorbed into the parton distributions functions, exactly as for the collinear $\alpha_S \log Q^2$ logarithms of perturbative QCD. There are two effects of this: the normal DGLAP evolution 
equations are slightly modified  --- the emmitted photon carries away some of the quark's momentum --- and a ``photon parton distribution" of the proton, $\gamma^p(x,Q^2)$, is generated. 
By correctly taking account of these QED effects through modified DGLAP evolution equations, 
we obtain a consistent procedure for dealing with this part of the overall electroweak 
correction in all hard-scattering processes involving initial-state hadrons (see for example 
\cite{earlyqed}).

Indeed, we might naively expect that the ${\cal O}(\alpha)$ contributions will be as numerically 
important as the 
${\cal O}(\alpha_S^2)$ NNLO QCD corrections. The only way to really find out 
is to perform
a full global parton distribution function analysis with QED corrections included, and to compare with the results of a standard QCD-only analysis. 
The first quantitative estimates of the effect on the evolution of parton
distribution functions was made in \cite{spiesberger}, and a recent 
investigation was made in \cite{Roth}. In fact the effect is found to be  small
over the bulk of the $x$ range compared with the effects of including  NNLO QCD contributions in the evolution,
since  even though $\alpha_S^3$ is similar in size to $\alpha$, the LO
QED evolution has none of the large logarithms that accumulate at higher
orders in the QCD corrections. Furthermore, for obvious reasons the gluon evolution 
is largely unaffected by the QED corrections.

A deficiency of previous investigations is that they tend to start with a set of standard 
partons, obtained from a QCD-only global analysis, and evolve 
upwards with QED effects switched on, rather than attempting to consistently determine 
a completely new set of QED-corrected 
partons from an overall  best fit to data. We will take this further step in this paper.
Although, as we shall see, the QED corrections have only a very small effect on the evolution
of quarks and gluons, they do have two interesting side effects. First, 
they necessarily lead to isospin violation, i.e. 
$u^p \neq d^n$, since the two quark flavours evolve differently when QED effects are included 
(unlike gluons, photons are not flavour blind). This is relevant to the 
NuTeV measurement of $\sin^2\theta_W$ from neutrino- and antineutrino-nucleus scattering, 
see for example \cite{NuTevtheta} and \cite{MRSTerror2}.
Second, the photon parton distribution may be large enough to be measureable in $ep$ collisions at HERA, by Compton scattering at wide angle off the electron beam.

In this paper we first discuss the QED-modified DGLAP equations and the form of the starting
distributions at $Q_0$.  We then, in Section~4,  
obtain numerical results for the resulting set of parton distributions within the framework of the standard MRST NLO and NNLO global
analysis.\footnote{Preliminary results from this study have been presented in Ref.~\cite{MRSTDIS04}.}
 In Section~\ref{sec:measuring} 
we discuss how the photon parton distribution may be 
experimentally measured. 

\section{DGLAP formalism including QED effects}

The factorization of the QED-induced collinear divergences leads to 
QED-corrected evolution equations for the parton distributions of the proton. These are (at leading order in both $\alpha_S$
and $\alpha$)
 \bea 
 {\partial q_i(x,\mu^2) \over \partial \log \mu^2} &=& {\alpha_S\over 2\pi}
\int_x^1 \frac{dy}{y} \Big\{
    P_{q q}(y)\; q_i(\frac{x}{y},\mu^2)
     +  P_{q g}(y)\; g(\frac{x}{y},\mu^2)
    \Big\} \nonumber \\
&  + &
   {\alpha\over 2\pi} \int_x^1 \frac{dy}{y} \Big\{
    \tilde{P}_{q q}(y)\; e_i^2 q_i(\frac{x}{y},\mu^2)  +  P_{q \gamma}(y)\;
e_i^2 \gamma(\frac{x}{y},\mu^2)         \Big\}  \nonumber \\
{\partial g(x,\mu^2) \over \partial \log \mu^2} &=& {\alpha_S\over 2
\pi} \int_x^1 \frac{dy}{y} \Big\{
    P_{g q}(y)\; \sum_j q_j(\frac{x}{y},\mu^2) 
 + 
    P_{g g}(y)\; g(\frac{x}{y},\mu^2)
    \Big\} \nonumber \\
   {\partial \gamma(x,\mu^2) \over \partial \log \mu^2}
& =   & {\alpha
\over 2\pi} \int_x^1 \frac{dy}{y} 
   \Big\{ P_{\gamma q}(y)\; \sum_j e_j^2\; q_j(\frac{x}{y},\mu^2) 
+ 
  P_{\gamma \gamma}(y)\; \gamma(\frac{x}{y},\mu^2) \Big\} \; ,
\eea
where
\bea
\tilde{P}_{qq} = C_F^{-1} P_{qq}, & &   P_{\gamma q} = 
C_F^{-1} P_{g q}, \nonumber \\
P_{q\gamma} = T_R^{-1} P_{q g} , & &  P_{\gamma \gamma} = - 
\frac{2}{3}\; \sum_i e_i^2\; \delta(1-y)
\eea
and momentum is conserved:
\beq
  \int_0^1 dx\;  x\; \Big\{\sum_i q_i(x,\mu^2) + g(x,\mu^2) + \gamma(x,\mu^2)
     \Big\}  = 1 \; .
\eeq
Note that, in principle, we could introduce {\it different} factorisation scales for the QCD 
and QED collinear divergence subtraction,
thus $q(x,\mu^2_{F{\rm (QCD)}}, \mu^2_{F{\rm (QED)}} )$ etc. with separate DGLAP equations 
for evolution with respect to 
each scale, but this is an unnecessary extra complication that we will ignore
and indeed, as is conventional, we will use $\mu_F^2=Q^2$ for DIS processes.

With the above formalism, it is in principle straightforward to repeat the global NLO or NNLO (in pQCD) fit.  However there is a complication because now we must allow for isospin symmetry 
breaking in all the distributions, that is
$\gamma^p \neq \gamma^n \Rightarrow q^p \neq q^n  \Rightarrow g^p \neq g^n$. 
This makes
the evolution and fitting significantly more complex, and potentially more 
than doubles the number of parameters in the fit, a signficant fraction of which will
not be at all well determined.

Therefore we adopt a simpler approximation which nevertheless contains the 
essential physics. Since it turns out that the dominant effect of the 
QED corrections is the radiation of photons off high-$x$ quarks we will 
assume that the isospin-violating effects at the 
starting scale $Q_0^2$ are confined to the valence quarks only.

Momentum conservation now reads
\bea
\int_0^1 dx \; x (\upv + \dpv + \gamma^p + S + g ) &=& 1 \nonumber \\
\int_0^1 dx \; x (\unv + \dnv + \gamma^n + S + g ) &=& 1  \; ,
\label{momcon1}
\eea
where we have  assumed that at $Q_0^2$, the sea quarks and gluon are isospin 
symmetric,
i.e. $S^p = S^n = S$, $g^p = g^n = g$. This symmetry is {\it not} preserved 
by evolution, but is only violated very weakly.

\section{The starting distributions}

We next assume that the photon distribution at $Q_0^2$ is that obtained by 
one-photon
emission off valence (constituent) quarks in the leading-logarithm 
approximation.
This is just a model, of course, but as long as these distributions are 
${\cal O}(\alpha)$
compared to the starting quark and gluon distributions, then they have a 
negligible effect
on the quark and gluon evolution. Thus we take photon starting distributions
of the form
\bea
\gamma^p(x,Q_0^2) & = & \frac{\alpha}{2\pi} \left[ \frac{4}{9} 
\log\left({Q_0^2\over m_u^2}\right) u_0(x) + \frac{1}{9} \log\left({Q_0^2
\over m_d^2}\right) d_0(x)\right] \otimes  {1+(1-x)^2 \over x}
 \nonumber \\
\gamma^n(x,Q_0^2) & = & \frac{\alpha}{2\pi} \left[ \frac{4}{9} \log
\left({Q_0^2\over m_u^2}\right)  d_0(x) + \frac{1}{9} 
\log\left({Q_0^2\over m_d^2}\right) u_0(x) \right] \otimes {1+(1-x)^2 \over x}
\label{gammastarting}
\eea
where $u_0$ and $d_0$ are `valence-like' distributions of the proton that satisfy
\beq
\int_0^1 dx\; u_0 = 2 \int_0^1 dx\; d_0 = 2\; , \qquad  \int_0^1 dx\; 
x(u_0+d_0) = 0.5 \; .
\eeq
The following functions have the required properties:\footnote{These model distributions are 
simply used to determine the starting distributions of the photon. The global analysis determines
the precise forms of $u_V$ and $d_V$ at $Q_0^2$.}
\beq
x u_0(x) = 1.273 \sqrt{x} (1+6.463 x)(1-x)^3 \; , \qquad
     x d_0(x) = 0.775 \sqrt{x}(1+6.463 x) (1-x)^4 \; .
\eeq
Next, we need a model of isospin-violating $u_V$ and $d_V$ starting 
distributions. We assume that the difference $\dnv-\upv$ is described by a 
numerically 
small function $f(x)$, whose zeroth moment vanishes to preserve the valence 
quark number, and whose first moment
is such that momentum is conserved at $Q_0^2$. Given that we would expect 
$f(x)$
to have valence-like shape as $x\to 0$ and $1$, a convenient choice is 
$f(x) = \epsilon \left(
\upv(x,Q_0^2) - 2\dpv(x,Q_0^2) \right)$ where 
$\epsilon$ is determined by momentum conservation. Thus
\bea
\dnv - \upv = 2 (\dpv - \unv)  &=&  \epsilon (\upv - 2 \dpv) \nonumber \\
\Rightarrow \quad \dnv &=& (1+\epsilon) \upv - 2 \epsilon \dpv \nonumber \\
\mbox{and} \quad \unv &=& (1+\epsilon) \dpv - \frac{1}{2} \epsilon \upv 
\label{nstarting}
\eea
where the first equality is assumed due to approximately twice as many photons being radiated
from $u^p$ as $u^n$ and vice-versa for the $d$ distributions.
Taking the difference of the two equations in Eq.~(\ref{momcon1}) 
at $Q_0^2$ gives
\beq
\int_0^1 dx \; x (\upv + \dpv -\dnv -\unv ) = 
 \int_0^1 dx \; x (\gamma^p - \gamma^n  ) 
\label{momcon2}
\eeq
and substituting for the neutron distributions  from (\ref{nstarting}) 
allows $\epsilon$ to be determined:
\beq
\epsilon = 2 \; {\int_0^1 dx \; x (\gamma^n - \gamma^p  )
\over \int_0^1 dx \; x (\upv - 2\dpv) } \; .
\label{epsilonequals}
\eeq
For the particular model for $\gamma^{p,n}(x,Q_0^2)$ introduced above, it is 
straightforward to calculate\footnote{We take $\alpha^{-1} = 137$, current quark masses $m_u= 6$~MeV, 
$m_d = 10$~MeV, and  $Q_0^2 =1$~GeV$^2$.} the numerator in (\ref{epsilonequals}):
\bea 
\int_0^1 dx \; x (\gamma^p - \gamma^n  ) & = &  \frac{\alpha}{2\pi} 
\left[ \frac{4}{9} \log\left({Q_0^2\over m_u^2}\right)\; 0.3573 + \frac{1}{9} 
\log\left({Q_0^2\over m_d^2}\right) \; 0.1427 \right] \times \frac{4}{3}
\nonumber \\
& - &
\frac{\alpha}{2\pi} \left[ \frac{4}{9} \log\left({Q_0^2\over m_u^2}\right)\; 
0.1427  + \frac{1}{9} \log\left({Q_0^2\over m_d^2}\right) \; 0.3573 \right] 
\times  \frac{4}{3} \nonumber \\
&=& \frac{\alpha}{2\pi}\; \frac{4}{27}\; 0.2146\; \left[  4\;
\log\left({Q_0^2\over m_u^2}\right)\; -
\; \log\left({Q_0^2\over m_d^2}\right)  \right]\ = \ 0.00117\; .
\label{numerator}
\eea
The denominator in (\ref{epsilonequals}) is just the  momentum fraction
carried by the valence up quarks minus twice the momentum fraction carried by 
the valence down quarks in the proton at the starting scale. For the 
 partons obtained in the new global (NLO pQCD) fit described below, this 
difference is $0.0746$, and substituting  gives $\epsilon = 0.0325$.

\begin{figure}[htb]
\begin{center}
\mbox{\epsfig{figure=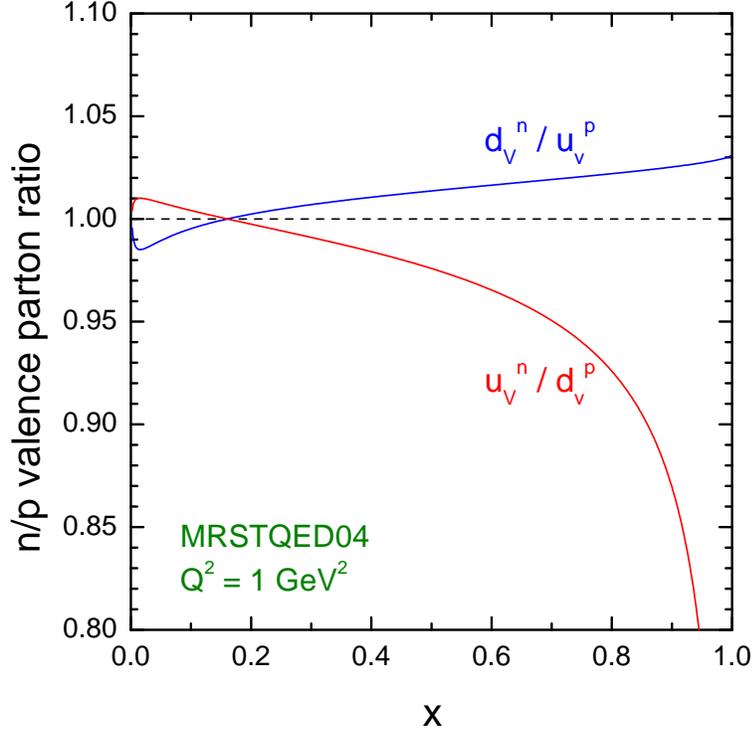,width=12cm}}
\vspace*{-5.5cm} \caption{\label{iso1} The ratio of valence quarks in the neutron and 
proton at the starting scale, $Q_0^2 = 1$~GeV$^2$, in the NLO global analysis, incorporating the isospin violation described 
by Eq.~(\ref{nstarting}).}
\end{center}
\end{figure}

Fig.~\ref{iso1} shows the ratio of the starting distributions of the neutron 
and the proton valence quarks, i.e. $\dnv/\upv$ and $\unv/\dpv$, for this 
value of $\epsilon$.  The deviation of these 
ratios from unity signals isospin violation in the starting distributions. 
We see that the result is as expected, with fewer high-$x$ up-quarks in the 
proton than down-quarks in the neutron due to increased radiation of photons. 
Similarly we see the expected excess of down-quarks in the proton compared to 
up-quarks in the neutron.  

It would be possible to devise other physically motivated models for 
the differences between $\upv(x,Q_0^2)$ and $\dnv(x,Q_0^2)$ and between  
$\dpv(x,Q_0^2)$ and $\unv(x,Q_0^2)$, for example we could estimate the change in
a quark distribution between scales $m_q^2$ and $Q_0^2$ due to QED evolution to be
\beq
\Delta q(x,Q_0^2) = {\alpha\over 2\pi} \int_x^1 \frac{dy}{y} 
    \tilde{P}_{q q}(y)\; e_q^2 \; q(\frac{x}{y},Q_0^2) \; \log(Q_0^2/m_q^2) \; ,
\label{deltaq}
\eeq
and make the differences between the input quarks for the proton and the 
neutron to be consistent with this. The momentum carried by the photon 
in the proton and neutron could then be determined by the momentum lost by
each quark due to this contribution. However, in practice this results in 
distributions and asymmetries which are very similar to those in our model,
with the essential features being identical. The
results are actually much more sensitive to issues such as the choice 
of the values of the quark masses. 
   
\section{Global analysis including QED effects}

Having defined our procedure for obtaining the QED contribution to the 
input partons, the strategy for the fitting procedure is then to

\begin{itemize}
\item[(i)] calculate the starting distributions $\gamma^p(x,Q_0^2)$ and  
$\gamma^n(x,Q_0^2)$;
\item[(ii)] parametrise the proton's quark and gluon distributions 
at $Q_0^2$  in the usual (MRST) way;
\item[(iii)] compute $\epsilon$ using Eq.~(\ref{epsilonequals}); 
\item[(iv)] calculate the neutron starting quark and gluon distributions  at 
$Q_0^2$ by assuming  isospin symmetry for sea quarks and gluons, and 
isospin-violating valence distributions given by Eq.~(\ref{nstarting});
\item[(v)] perform the global fit, using separate DGLAP equations for the 
proton and neutron partons. 
\end{itemize}
We have performed fits at both NLO and NNLO, where the NNLO fit uses the 
recently calculated exact NNLO splitting functions \cite{MVVns,MVVs}. We use the same
input data\footnote{Note that by using the identical set of data as used in the standard fit we are 
implicitly assuming 
that no QED corrections corresponding to photon emission off incoming quark lines have been applied.} 
 as in the recent MRST2004 study of Ref.~\cite{MRST2004}.
In both cases the QED corrections do not alter the fit quality in any 
significant way. 
For the NLO fit with QED corrections the $\chi^2$ is actually $\sim 15$ higher 
than that for the standard NLO fit. This increase comes from two sources. 
The very small amount of 
momentum carried by the photon is effectively taken from the gluon -- the
size of the input quarks being very well fixed by the data. This conflicts with our 
usual findings that at NLO the gluon would actually like more momentum 
both at high $x$, in order to fit the jet data, and at moderate $x$ ($\sim 
0.1-0.01$), in order to fit the slope of the HERA and NMC structure function data. In order to 
compensate for this loss of gluon the value of $\alpha_S(M _Z^2)$ increases 
very slightly, by about $0.0002$, but the fit to the H1 data is still worse 
by about $8-10$ units of $\chi^2$. Also, the new mechanism of photon radiation, 
preferentially from high-$x$ up-quarks, tends to make $F_2^p(x,Q^2)$ fall more 
quickly with $Q^2$ at high-$x$, and this is effect is increased by the slight
increase in $\alpha_S(M_Z^2)$. This makes the fit to the BCDMS proton 
structure function data $10$ units worse, as this data set prefers a slower fall off
with increasing $Q^2$. The fit to all other sources of data is 
actually about $5$ units better than the standard NLO fit, with the fit to
deuterium data being very slightly improved in general. The overall increase 
in $\chi^2$, whilst being significant, cannot be taken as evidence that QED 
effects should be ignored. They are most certainly present. Rather it 
highlights the minor shortcomings in the NLO QCD fit, most particularly the 
tensions between the gluon and $\alpha_S$.

This conclusion is borne out by the result of the NNLO fit with QED 
corrections. In this case the $\chi^2$ is lower than for the standard NNLO
fit, albeit only by 3 units. At NNLO the tensions between the gluon and
$\alpha_S(M_Z^2)$ are much reduced, and the QED corrections do not cause 
even minor problems in this respect. Indeed, the value of 
$\alpha_S(M_Z^2)$ is essentially unchanged. The small improvement in $\chi^2$ 
is due to slight improvements in the descriptions of the CCFR $F_3(x,Q^2)$ \cite{CCFR3}, 
BCDMS $F_2^d(x,Q^2)$ \cite{BCDMSd} and E866 Drell-Yan hydrogen/deuterium ratio data \cite{E866}, 
all of which are sensitive to the isospin
violation induced by the QED evolution. In the context of the overall fit, however, these improvements are too small to draw any definite conclusions.   

We can also perform the fit making a different assumption about the 
light-quark masses. In particular, we can take the extreme case of 
constituent-type quark masses 
of $300$~MeV for both the up and down quarks. From Eq.~(\ref{gammastarting})
 it is easy to see that 
this decreases the momentum carried by the photon at input very 
significantly, and consequently also decreases the input isospin asymmetry.
In this case $\epsilon = 0.0074$ at $Q_0^2$, to be compared with 
 $\epsilon = 0.0325$ for the previous (current quark mass) fit. 
However, the loss of gluon momentum is still generated by the subsequent 
evolution, and so this procedure 
only improves the quality of the NLO fit very slightly indeed, 
giving a $\chi^2$ of only $\sim 2$  lower than the previous fit. At NNLO 
there is also an improvement compared to the current quark mass prescription, 
but even smaller than at NLO. Hence, there is essentially no 
evidence from the global fit whether current quark mases or constituent
quark masses are preferred. We will return to this distinction between quark masses later.
\begin{figure}[htb]
\begin{center}
\mbox{\epsfig{figure=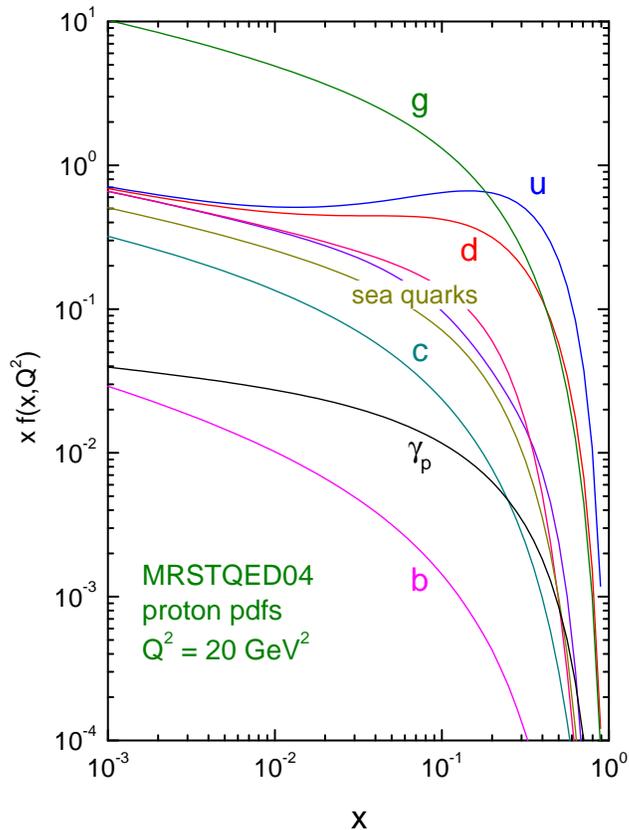,width=10cm}}
\vspace*{-2cm} 
\caption{\label{iso2} The parton distributions in the proton 
at $Q^2=20\;\GeV^2$ obtained from the NLO pQCD $+$ LO QED global fit. The curves for the sea quarks correspond to the 
$\bar u$, $\bar d$, $s$, $c$ and $b$ distributions.}
\end{center}
\end{figure}

\begin{figure}[htb]
\begin{center}
\mbox{\epsfig{figure=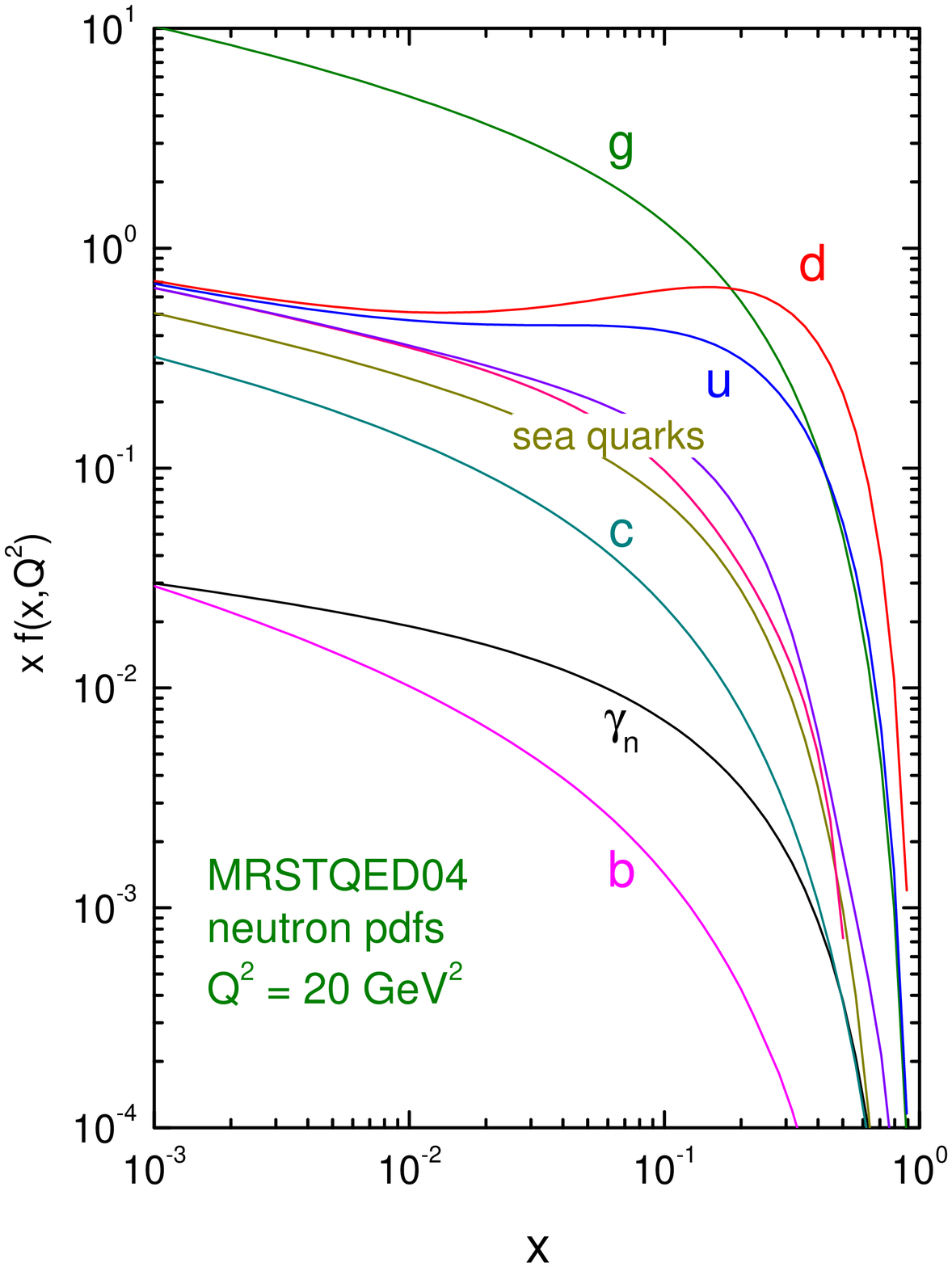,width=10cm}}
\vspace*{-2cm} 
\caption{\label{iso3} The parton distributions in the neutron 
at $Q^2=20\;\GeV^2$ obtained from the NLO pQCD $+$ LO QED global fit. The curves 
for the sea quarks correspond to the 
$\bar u$, $\bar d$, $s$, $c$ and $b$ distributions.}
\end{center}
\end{figure}

The parton distributions generated in the fit with the current quark masses, 
which we will treat as the default fit,\footnote{We believe that current quark masses are 
more appropriate than constituent quark masses because photon radiation is an entirely
perturbative QED effect which should not be sensitive to the 
strong scale or mass of hadrons. The default parton sets, which we denote by MRSTQED04, can be found at 
http://durpdg.dur.ac.uk/hepdata/mrs.html} are shown in Fig.~\ref{iso2}. The quark 
and gluon distributions are all extremely similar to the standard MRST parton 
distributions, but it is interesting to note the features of the new photon 
distribution. At $Q^2=20\; \GeV^2$ it is larger than the $b$-quark distribution, 
but this is because the $b$ quark is being probed not far above the
scale ($Q^2 = m_b^2$) where it turns 
on from zero at NLO. However, the photon distribution is larger than the 
sea quarks at the highest values of $x$. This is presumably because it
is generated directly from the radiation off high-$x$ valence quarks, whereas 
the sea quarks first branch into gluons which then subsequently produce 
sea quarks at even smaller momentum fractions.  The photon has a similar 
shape to the sea quarks at small $x$ since it is generated via the splitting 
function $P_{q\gamma}$ which gives a contribution proportional to the size of 
the quarks at the smallest $x$ values. In Fig.~\ref{iso3} we show the corresponding figure 
for the parton distributions in the neutron. The quarks and gluon are almost 
indistinguishable from those in the proton, once one interchanges up- 
and down-quark distributions, but the photon distribution is 
smaller at large $x$, as we would expect from the 
decreased charge squared of the dominant valence quarks. The photon distributions
of the proton and neutron become similar at very small $x$, reflecting the charge symmetry of the 
small-$x$ sea quarks.       
\begin{figure}[htb]
\begin{center}
\mbox{\epsfig{figure=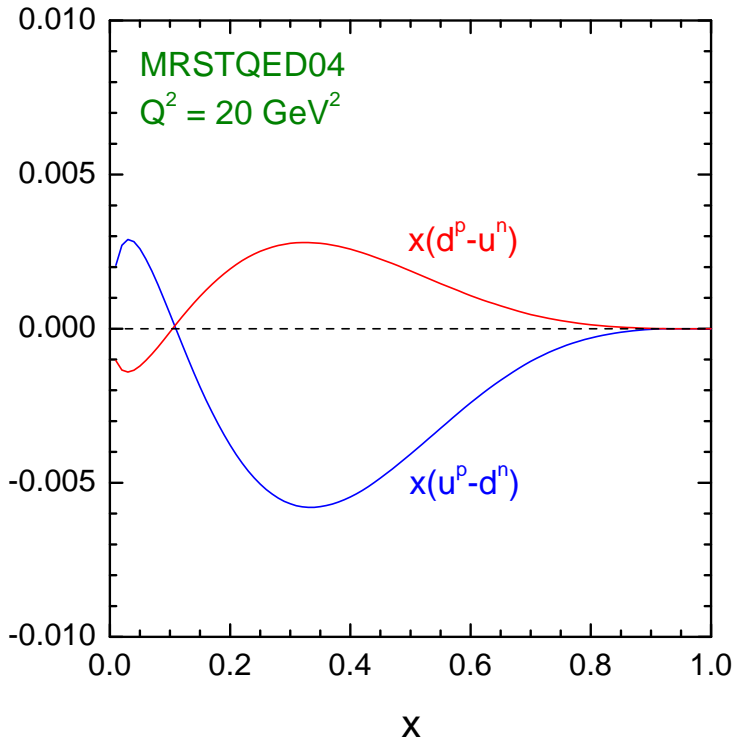,width=12cm}}
\vspace*{-5.5cm} \caption{\label{iso4} The difference between the isospin 
exchanged valence quarks in the proton and the neutron  
at $Q^2=20\;\GeV^2$.}
\end{center}
\end{figure}
In Fig.~\ref{iso4} we plot the valence-quark differences $x(\dpv-\unv)$ and $x(\upv-\dnv)$ at 
$Q^2=20\; \GeV^2$. This figure illustrates the violation of isospin symmetry in 
the momentum carried by the valence quarks particularly clearly.   
As mentioned earlier, this has important implications for the anomaly in the 
measurement of $\sin^2\theta_W$ reported by NuTeV \cite{NuTevtheta}. The quantity 
measured, up to corrections due to cuts \cite{NuTevtheta,NuTeVtheta2}, 
by NuTeV is
\beq
 R^-=\frac{\sigma^{\nu}_{\rm NC}
-\sigma^{\bar\nu}_{\rm NC}}{\sigma^{\nu}_{\rm CC}
-\sigma^{\bar\nu}_{\rm CC}}. \label{eq:rminus}
\eeq
In the simplest approximation, i.e. assuming an isoscalar target, 
no isospin violation and equal strange and anti-strange distributions, 
this ratio is given by
\beq
R^-\approx \frac{1}{2}-\sin^2 \theta_W,\label{eq:rminusn}
\label{eq:simple}
\eeq 
and so the measurement gives a direct determination of $\sin^2 \theta_W$. 
NuTeV find $\sin^2 \theta_W=0.2277 \pm 
0.0013\; ({\rm stat.}) \pm 0.0009\; ({\rm syst.})$ \cite{NuTevtheta}, compared to 
the global average of $0.2227 \pm 0.0004$,
that is, roughly a $3\sigma$ discrepancy. However, if one allows for isospin 
violation then the simple expression (\ref{eq:simple})
becomes modified to
\beq 
R^-=\frac{1}{2}-\sin^2 \theta_W +(1-\frac{7}{3}
\sin^2 \theta_W) \frac{[\delta U_{\rm v}] -[\delta D_{\rm v}]}{2[V^-]}
\label{eq:rminusi},
\eeq
where
\beq
 [\delta U_{\rm v}] = \int_0^1 dx \,x(u^p_{\rm v}(x) - d^n_{\rm v}(x)), 
\qquad\qquad  
[\delta D_{\rm v}] = \int_0^1 dx \,x(d^p_{\rm v}(x) - u^n_{\rm v}(x)),
\label{eq:momdefs}
\eeq  
and $[V^-]$ is the overall momentum fraction carried by the valence quarks.

In the extraction of the value of $\sin^2\theta_W$, a correction is made to take account of the
electroweak corrections to the cross section. These corrections contain the 
collinear singularities absorbed into the QED evolution of partons, and so must
not be double-counted. The most recent calculations of these corrections
\cite{ewnutev} {\it do} factor out the collinear singularities, and are thus
designed to be used with QED-corrected partons. In the electroweak 
corrections used by NuTeV \cite{ewnutevorig} the collinear singularities were
regularised by giving the quarks a mass of $xm_p$, which is rather large for 
the most important region of high $x$, and effectively allows less radiation 
from high $x$ than low $x$, minimising the isospin-violation effect of QED 
radiation. Hence, this procedure should be updated, but there is certainly
minimal double counting employed by using our QED corrected partons
even in this case.   
   
Since the isospin violation generated by the QED evolution is precisely
such as to remove more momentum from up-quark distributions than down-quark 
distributions, it clearly works in the right direction to reduce the NuTeV
anomaly. The effect is also $Q^2$-dependent, since the quantities in
Eq.~(\ref{eq:momdefs}) have a non-zero anomalous dimension. At $Q^2=2\;\GeV^2$ we have 
$[\delta U_{\rm v}]=-0.002271$, $[\delta D_{\rm v}]=0.001124$ and 
$[V^-]=0.4428$, leading to a change in the measured value of $\sin^2 \theta_W$ of 
$- 0.0018$, i.e. a little more than $1\sigma$ of the total discrepancy 
is removed. It is not obvious how this result will change with
$Q^2$, since as $Q^2$ increases all the valence distributions evolve to
smaller $x$ and the momentum carried by each will decrease. However, the 
isospin-violating component of the evolution is present, and so we might expect
an increase in the effect. Indeed, at $Q^2=20\;\GeV^2$ we find 
$[\delta U_{\rm v}]=-0.002095$, $[\delta D_{\rm v}]=0.001005$ and 
$[V^-]=0.3501$ leading to a change in measured value of $\sin^2 \theta_W$
of $-0.0021$. This general trend continues with increasing $Q^2$,
reaching $\Delta \sin^2 \theta_W = -0.0029$ at $Q^2=20000\; \GeV^2$. These results are
in remarkable agreement with our previous analysis of isospin-violating effects in parton distributions based on the Lagrange Multiplier method, see Section~5.4 of Ref.~\cite{MRSTerror2}.
There we found a shift of $\delta R^-_{\rm iso} = -0.002$, with  90\% confidence level limits
of $- 0.007 < \delta R^-_{\rm iso} < +0.007$, comfortably more than needed to explain
the NuTeV anomaly. 

Hence we conclude that the QED contribution to isospin violation in the 
valence quarks has a significant effect in reducing the value of $\sin^2\theta_W$
as measured by NuTeV.  We note also that 
the naive results quoted above need to be corrected for the acceptance cuts made on the 
data. Functions for convolving with the parton distributions to take these acceptance effects
into account are provided in \cite{NuTeVtheta2}. However these do not contain any
$Q^2$-dependence, despite accounting in principle for the momentum fraction
carried by the valence quarks, which is certainly a scale-dependent
quantity. Hence we can only estimate that the corrections may reduce the 
observed effect by $10-20\%$, see the discussion in Ref.~\cite{MRSTerror2}. We also note that the quoted results can be 
diminished by a factor of up to about 4 if constituent quark masses of
$300$~MeV are used instead of current masses -- however this option is neither
experimentally nor theoretically favoured.

\section{Measuring the photon parton distribution, $\gamma(x,Q^2)$}
\label{sec:measuring}

The photon parton distributions of the proton and neutron, $\gamma^p$ and $\gamma^n$, 
are a direct and inescapable consequence of introducing
QED contributions into the DGLAP equations. It is therefore interesting to speculate how
they could be measured directly in experiment. In particular, such a measurement would test our model
assumption for the starting distributions $\gamma(x,Q_0^2)$ given in Eq.~(\ref{gammastarting}). 

\begin{figure}[htb]
\begin{center}
\mbox{\epsfig{figure=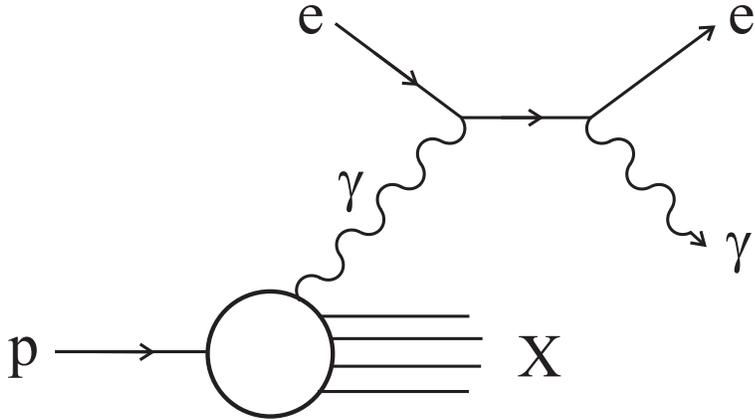,width=15cm}}
\caption{\label{fig:compton}  Schematic diagram for the deep inelastic scattering
process $ep \to e\gamma X$, which displays the convolution
of $\gamma^p$ and $\hat{\sigma}(e\gamma \to e\gamma)$
of (\ref{sigamma1}). Besides the $s$-channel diagram for $e\gamma \to e\gamma$
that is shown, there is also a contribution from the diagram
with a virtual $u$-channel electron.}
\end{center}
\end{figure}
The most direct measurement of the photon distribution in the proton 
would appear to be wide-angle scattering of the photon by a charged lepton beam, thus  $ep \to e \gamma X$ where the final state electron and photon are produced with equal and opposite large transverse momentum. 
The subprocess is then simply QED Compton scattering, $e \gamma \to e \gamma$, and the cross section is obtained by convoluting this subprocess cross section with $\gamma^p$, see Fig.~\ref{fig:compton}, 
\beq
\sigma (ep \to e \gamma X) = \int dx^\gamma \; \gamma^p(x^\gamma,\mu^2)\; \hat\sigma(e\gamma \to e \gamma) \; ,
\label{sigamma1}
\eeq
where $\mu$ is the factorisation scale.
If the photon is produced with transverse energy $E_T^\gamma $ and pseudorapidity $\eta^\gamma$ in the
HERA laboratory frame, then simple kinematics gives 
\beq
x^\gamma = {  E_T^\gamma   E_e \exp(\eta^\gamma)  \over  2 E_p E_e -  E_T^\gamma   E_p \exp(-\eta^\gamma)   } \; ,
\label{sigamma2}
\eeq
where $E_e$ and $E_p$ are the energies of the electron and proton beams respectively.

The ZEUS collaboration \cite{ZEUS}  has recently published a measurement of this
cross section:
\beq
\sigma(ep \to e \gamma X)\; =\; 5.64\; \pm\; 0.58\; \mbox{(stat.)}\; {\small{+0.47 \atop -0.72}}\; \mbox{(syst.)}~\mbox{pb.}
\eeq
in electron-proton collisions\footnote{In fact, the data sample corresponds to a mix of electron and positron beams, but obviously the corresponding theoretical predictions are identical.} with $\sqrt{s} = 300$ and $318$~GeV. The final state cuts are
\bea
5 < E_T^\gamma < 10\ \mbox{GeV}\; , &\qquad& -0.7 < \eta^\gamma < 0.9\; , \nonumber \\
Q^2 > 35\ \mbox{GeV}^2\; , &\qquad& E_{e'} > 10\ \mbox{GeV}\; , \qquad {139.8}^\circ      
< \theta_{e'} < {171.8}^\circ \; .
\label{ZEUScuts}
\eea
It is noted in \cite{ZEUS} that neither PYTHIA nor HERWIG can explain the observed rate
(underestimating the measured cross section by factors of 2 and 8 respectively) or
(all of) the kinematic distributions in $E_T^\gamma$, $\eta^\gamma$ and $Q^2$.

Using the proton's photon parton distribution obtained in the previous section and using the
same cuts as in (\ref{ZEUScuts}), we find
\beq
\sigma(ep \to e \gamma X)\; =\; 6.2\; \pm\; 1.2\ \mbox{pb.} 
\label{eq:prediction}
\eeq
where the error corresponds to varying the factorisation scale in the range
$E_T^\gamma/2 < \mu < 2 E_T^\gamma$ with $\mu = E_T^\gamma$ taken as the central value. The fact that this \lq parameter-free' prediction
agrees well with the experimental data lends strong support to our analysis and, in particular, to our choice of current quark masses in defining the initial photon distribution. As already pointed out, the photon distribution obtained with constituent quark masses is smaller, and in fact
reduces the theoretical prediction of (\ref{eq:prediction}) to 3.6~pb, in disagreement with the measured value. 
It would
be interesting to extend the ZEUS analysis to make a direct measurement of $\gamma^p(x^\gamma,Q^2)$ 
as a function
of $x^\gamma$, using Eqs.~(\ref{sigamma1},\ref{sigamma2}). 
In the measurement reported in \cite{ZEUS}, $x^\gamma$ is sampled
in a fairly  narrow range centred on $x^\gamma \simeq 0.005$.

\section*{Acknowledgements}

We would like to thank Johannes Bl\"umlein and Jon Butterworth for useful discussions. RST would like to thank
the Royal Society for the award of a University Research Fellowship. RGR and ADM
would both like to thank the Leverhulme Trust for the award of an Emeritus
Fellowship. The IPPP gratefully acknowledges financial support from the UK
Particle Physics and Astronomy Research Council.


\end{document}